# Designing an Optimal Portfolio for Iran's Stock Market with Genetic Algorithm using Neural Network Prediction of Risk and Return Stocks

Masoud Fekri [1,*], Babak Barazandeh [2]

[1] School of Industrial Engineering, Iran University of Science and Technology, Tehran 16846-13114
[2] Department of Industrial and Systems Engineering, University of Southern California, Los Angeles, CA 90089, USA
( * corresponding author: masoud_fekri@ind.ius.ac.ir)

## Abstract
Optimal capital allocation between different assets is an important financial problem, which is generally framed as the portfolio optimization problem. General models include the single-period and multi-period cases. The traditional Mean-Variance model introduced by Harry Markowitz has been the basis of many models used to solve the portfolio optimization problem. The overall goal is to achieve the highest return and lowest risk in portfolio optimization problems. In this paper, we will present an optimal portfolio based the Markowitz Mean-Variance-Skewness with weight constraints model for short-term investment opportunities in Iran's stock market. We will use a neural network based predictor to predict the stock returns and measure the risk of stocks based on the prediction errors in the neural network. We will perform a series of experiments on our portfolio optimization model with the real data from Iran's stock market indices including Bank, Insurance, Investment, Petroleum Products and Chemicals indices. Finally, 8 different portfolios with low, medium and high risks for different type of investors (risk-averse or risk taker) using genetic algorithm will be designed and analyzed.

## A. Motivation and Significance
Financial optimization, including asset allocation and risk management is an attractive area in uncertain decision-making. In addition to the asset allocation, the allocation percentage of the total portfolio's value to the portfolio component, risk measurement and management of investment tools, and the creation or maintenance of a portfolio with a risk-return profile are other financial optimization issues. Although asset allocation and risk management are the two main parts of financial optimization, risk management is a fundamental issue for the management of financial institutions since 1990. Elton et al [1]. So in this paper, we will be proposed three portfolios with low, medium and high risks for different type of investors for risk management of portfolios.

## B. Introduction
Portfolio optimization is a method that quantifies the selection of different investment options to reach the maximum return in a risk level. There are several approaches with different objective functions to design an optimal portfolio. Financial institutions must consider the uncertainties over time, legal restrictions and other policies to design their organization's strategies such that the combination of desirable properties meets their goals.
Investing in the stocks of the Stock Exchange is one of the lucrative options in the capital market. The stock market is a non-linear and chaotic system that is affected by political, economic and psychological circumstances. In order to achieve long-term growth and sustainable economic activities such as providing stock liquidity, increased transparency of information, the possibility of picking up small capitals, application of the rules of corporate governance is a particularly important task of the Stock Exchange. In the early twentieth century, a group of professionals experienced in evaluating securities believed that a prediction for the stock price can be provided through the study and analysis of historical stock price changes. More scientific studies with emphasis on the identification of the behavior of stock prices tend to the stock price valuation models.

The theory of random walks was a start in determining the behavior of stock prices. Later, investigation of the features and structure of the capital market resulted in the efficient capital market hypothesis.

In efficient capital market, it is believed that stock prices reflect the current information on the stock and stock price changes don't follow a predictable pattern. Theories proposed prior to the 1980s were able to determine the stock prices' behavior in the market until the New York stock market events in 1987 sharply questioned the validity of the efficient market hypothesis and stochastic price models. Since 1990s, more specialists have favored a chaotic behavior with some sort of discipline and increasing effort has been done to design nonlinear models to predict stock prices.

One of the important issues raised in the capital markets is the optimization of the investment portfolio with respect to the risk and return of investment. Usually it is assumed that investors do not like risk and always seek to invest in properties that have the highest return and lowest risk. In other words, investors consider investment returns as a favorable factor to maximize and the variance of returns (risk) as undesirable factor to be minimized.

Iran's capital market whose main activity has been in the Tehran Stock Exchange (with a history of about 38years) and focused on the stock transactions, suffers from a fundamental weakness. Lack of specialists in the field has led to the creation of a stationary economic system with inefficient markets. Even though, the market has seen the return of 400 percent on the capital market which has imposed severe arbitrage positions. Some of the reasons for the weakness of the Iranian capital market include but not limited to, lack of financial engineering experts to estimate the impacts in stock price prediction, inefficiency in the stock market, which could negatively affect stocks' value, insufficiency and inefficiency of capital market regulation, rule-based system of banks in the financial system, non-utilization of the instruments or popular financial assets in global financial markets, lack of diversification of financial instruments on the stock exchange, lack of diversity in institutions and financial intermediaries, lack of mechanisms for participation of foreign investors in the securities market, insufficiency of tax laws, lack of official exchange for cash and futures trades of foreign currencies, lack of active management of investment and technical analysis.

Due to the weakness of the capital market in Iran, in this paper, we will provide a technical analysis for prediction of stock return trends in the stock market under an active investment management. Technical analysis is a method to predict market prices through the study of past market conditions. In this analysis by exploring changes and fluctuations in price and trading volume in addition to supply and demand situations, future trends of prices is predicted. This analysis is widely used in the foreign exchange, stock, securities, gold and other precious metals' markets. Some of the goals in modern technical analysis is to enhance the ability of analysts to read and understand the market behavior expectations, reducing dependence on visual analysis, determining supply and demand response in time, understanding the relationship between time, cost and process structure, early detection of violations of patterns and market trends, and finally to increase the ability to receive early warning signs.

Mean-variance model by Markowitz [2] presented an idea and a guide for modern portfolio theory. This model is used in two ways, either the investment risk is minimized in a fixed expected rate of return, or the return of investment is maximized for an acceptable level of risk. Many researchers after his proposed model tried to extend his work or propose new models. As an example, Konno [3] extended Mean-variance model to include skewness. Yoshimoto[4] modifies the Mean-variance model by adding transaction cost . Soleimani [5] proposed adding minimum transaction lots, cardinality constraint and market capitalization constraints simultaneously and used a genetic algorithm for solving it. Fernández[6] generalizes Markowitz Mean-Variance model to include cardinality and bounding constraints to insure investing in different number of assets and to enforce limitation on investment in each asset and uses neural network to solve the resulting optimization problem. Konno [7] proposed a new model that uses mean absolute deviation risk for portfolio optimization. Vercher[8] proposed a probabilistic model for the portfolio selection problem which is based on a multi-objective optimization model that are related to probabilistic mean-downside risk-skewness model.

## C. Literature Review

Portfolio selection problem is one of the classic problems of the financial world that was first introduced by Markowitz [9], which includes two main components, return and the risk. The main purpose of this problem is to maximize the expected return on a certain level of risk or minimize the expected risk on a certain level of return Elton et al. [1], Campbell et al. [10]. Studies that have been done for single-period models include several parts, which are discussed next.

### C.1. Markowitz Model

Bienstock [11] used the quadratic model of Markowitz and derived indices for 3897 different assets using branch-and-cut algorithm. Chang et al. [12] used the Mean-Variance model to find the efficient frontier by presenting three heuristic methods, genetic algorithms, tabu search and simulated annealing. In this paper, comparison criteria will be chosen to be the existence of cardinality limit. Goll and Kallsen [13] studied the expected logarithmic utility maximization problem in a semi martingale market model. León et al. [14] extended their methods according to conditions on the portfolio that are imposed by investor. These conditions may force infeasibility in mathematical models because of lack of experience or knowledge between set of circumstances. Their method is based on imposing some constraints in accordance with the preferences of decision-makers and ask the decision-makers about how reasonable these constraints are. Crama and Schyns [15] used simulating annealing (SA) for portfolio management. Their method is useful for determining the efficient frontier of Mean-Variance Model for mid-sized problems in reasonable time.

Ammar and Khalifa [16] applied convex quadratic approach in order to choose an optimum fuzzy portfolio. Yang [17] applied genetic algorithm (GA) along with a dynamic portfolio optimization system in order to improve the portfolio returns. In addition to GA, they have applied mean-variance elliptic methods. The information applied in this research includes the indices of total returns from six different Stock Exchanges (Canada, France, USA, Germany, Japan and UK). The information related to 60-month period were applied as historical data in order to determine stock weights. Finally, GA is proved to have higher return and lower risk compared to the other two methods. So was multistage GA compared to single stage GA. Applying cardinality constraints, Liu and Gao [18] presented an approach to solve portfolio optimization problem adequately, so that whenever a sum of money is invested in an asset, there should be a particular share. Convergence of Lagrange was applied in this research along with cardinality constraints as an inequality. The calculation results for 30 assets in Hong Kong Stock Exchange were investigated.

Deng et al. [19] studied the portfolio optimization problem based on Markowitz's model under cardinality constraints applying improved particle swarm optimization method. This problem was investigated over five Stock Exchanges including Dax 100, Hang Seng 31, Mikkei 220, S&P 100, and ETSE 100 from Mars 1992 to September 1997. In most cases, the performance of POS method is better than genetic, simulated annealing, and Tabu search algorithms. It was shown that, compared to the different methods, POS has an effective and strong performance, particularly in low-risk investments. Ma et al. [20] investigated cardinality constrained portfolio optimization problem over Shanghai and Shenzhen Stock Exchange based on hybrid differential evolution method. Results indicated the performance of their algorithm.

Castellano and Cerqueti investigated the portfolio optimization problem in presence of risky assets which are traded less frequently and contain lower liquidity. For a dynamic modeling of non-cash assets, a pure jump process was applied, which finally led to development of optimum portfolio. The theoretical models in this study were analyzed through Monte Carlo simulation which provides a useful financial prospect. [21]

A large number of models are appeared after the Markowitz's mean-variance model applying fundamental assumptions [22]. In all of these models that are now known as classic models, the expected portfolio return is obtained through a linear combination of the share of each stock in the portfolio and its expected return. The risk level is variable, but it is mostly considered to be based on some moment of average linear combination of time series for the returns on existing stocks in the portfolio. Although the classic models of portfolio selection are widely adopted, in many cases, their fundamental assumptions are in contrast to the real world. There exists return series that are not usually normally distributed and have skewness and kurtosis [23], [24]. In such cases, the variance (or standard deviation) of returns is not enough to measure stock risk [25]. Moreover, utilization of mean return as a measure of future return of the stocks has a slight filtering effect on the dynamic behavior of stock market. Inaccurate estimations of future returns affects the performance of models in short-term and can cause losses in such investments.

Predictability of stock markets is still an open problem in the financial theories. The efficient market hypothesis provides a theoretical framework on this problem. Empirical tests and studying past several decades of the stock market can be helpful in this area [26], [27]. In an efficient market, a stochastic model is applied for some prices of stocks; however, price disorders and predictable patterns such as serial correlations, calendar effect and even sport results can affect that model and its prices [28].

In this section, further research has been provided by researchers for the Markowitz model:

## C.2. Mean-Variance model with limited components

Fernández and Gómez [6] reported the efficiency of applying Hopfield neural networks in solving constrained Markowitz problem. In their problem, there are cardinality and boundary constraints that are based on the Markowitz model. They compared Hopfield neural network to metaheuristic algorithm such as genetic algorithms (GA), simulated annealing (SA) and Tabu search (TS).

## C.3. Mean-Variance-Skewness model

Yu et al. [29] introduced a new radial basis function (RBF) neural network in order to choose an optimum portfolio based on mean-variance-skewness model. According to this RBF neural network, an investor can choose a portfolio simultaneously in line with its risk preferences and mean-variance-skewness objectives.

### C.4. Mean-Absolute deviation model
Speranza [30] considers the portfolio risk as linear combination of mean-absolute deviation (MAD) and proceeded to resolve the problem by applying a combination of integer linear programming and other heuristic methods. The results were tested over 20 securities.

### C.5. Mean-Semi variance model
Huang [31] introduced mean-semi variance model to choose an optimum portfolio. His solution was based on GA. Yan et al. [32] applied a combination of particle swarm optimization and genetic algorithm in multi-period portfolio using semi-deviation as risk measure. They showed that a combination of particle swarm optimization method and GA is more efficient than the mere use of each one.

### C.6. Multiple criteria model
Parra et al. [33] considered fuzzy objective function and constraints for portfolio optimization. Their solution was based on goal programming (GP) and the problem objectives included risk, return, and liquidity. GP method is a simultaneous solution for multiple and conflicting objective measure [34]. Costa and Paiva [35] investigate the robust optimal portfolio selection problem with a tracking error where the expected return of risky and risk-free assets and the covariance matrix of risky assets are not accurately specified. They proved that the following two problems are equivalent with linear-matrix optimization problem.
First problem: finding a portfolio with the minimum probability of the worst case of tracking error variations.
Second problem: finding a portfolio with maximum probability of the worst expected performance of the objective Ida [36] applied Interval objective function coefficients in multiple objective programming problem in order to choose a portfolio.

### C.7. Portfolio selection through time series prediction methods
There are different methods for prediction of financial time series. These methods can be divided into four groups:
1) Fundamental analyses
2) Technical analyses
3) Prediction of time series through traditional models
4) Machine learning methods
Fundamental analyses are based on economic factors of the market, such as analyses of statistics, projects, supply and demand conditions, services, economic principles and financial analysis of companies.
Prediction of time series through traditional models aims to design linear prediction models in order to determine the trend of historical data. Linear models can be categorized in two groups: multi variate and single-variate regression models. Recently, some machine learning methods are used for time series prediction. These methods get a set of sample data for linear and none-linear algorithms in order to obtain a patter.
HellstrSom and HolmstrSom [43] showed that it is not easy to predict the stock price. The academic researchers in this field are divided into two categories: those who believe that it is possible to design some mechanisms to predict the market behavior and those who believe that the stock prices change randomly (unpredictably) according to efficient market hypothesis by Fama [44].

## D. Problem Definition
In this section, we define the general problem and other definitions.

### D.1. Preliminary Definitions
In this subsection, we provide the basics of calculating the return and risk that is associated with a portfolio, which is used in the neural network model.

*Neural Network*: In order to design an investment portfolio, the investment risk and return data of a particular time horizon is observed in order to compare the quality of portfolios. Moreover, a predictive portfolio selection model for short-term strategies of investment requires a reliable prediction method of future stock returns and related risks. We utilize the Autoregressive Neural Network with p parameters ($ARNN(p)$) to predict the future return of the stock. Therefore, to predict the stock return at time $t+1$, denoted by $\hat{R}_{t+1}$ the return data of $((R_{t-(p-1)}, ..., R_{t-1}, R_t))$ are used.

A portfolio represents a distribution of an M number of stocks where the weight of a stock represents the relative share of that stock in the portfolio. The expected return of a portfolio, $\bar{R}$, is nothing but the expected value of the return of the stocks weighted by their corresponding weights (denoted by $X_i \geq 0$ for $1 \leq i \leq M$). Therefore, $\bar{R} = \sum_{i=1}^{M} X_i \hat{R}_t$. Moreover, on the normality assumption of the additive noise of the prediction errors, and defining the portfolio risk as the error covariance, the portfolio optimization cost can be defined as follows:

$$\hat{J}(X) = \sum_{t=1}^{M} X_t^2 \sigma_t^2 + \sum_{i=1}^{M} \sum_{j=1}^{M} X_i X_j Cov(i,j)$$
(D.1)

Assuming that the time-series prediction errors has a normal distribution, the portfolio risk is calculated as the variance of a linear combination of errors:

$$\hat{V} = \sigma_t^2 = \sum_{i=1}^{M} \sum_{j=1}^{M} X_i X_j Cov(i,j) \qquad (D.2)$$

Where $\hat{V}$ and $X_i$ represent the total portfolio risk and weight of the *i*th stocks in the portfolio, respectively. Moreover, $Cov(i,j)$ is the covariance of the predicted errors of the *i*th and *j*th stocks, which represents the predicted mutual risk of them, and is calculated as:

$$Cov(i,j) = \frac{1}{N-1} \sum_{t=1}^{t=N} \varepsilon_{it} \varepsilon_{jt} \qquad (D.3)$$

Therefore, equation (D.1) can be re-written as:
So that the first sum represents the distribution of the projected risk of each stock on the portfolio risk, and the second sum represents the mutual predicted risk of the *i*th and *j*th stocks.

### D.2. Portfolio Optimization with Risk Measures
Portfolio optimization deals with capital allocation among several assets. Portfolio optimization is an important research area in the field of modern risk management. Generally, it is in the interest of an investor to achieve the maximum return with minimum risk possible in the stock portfolio. Nevertheless, the high return results in a higher risk, as well. While Markowitz's Mean-Variance model, which leads to a quadratic programming, became popular in the recent years, it is mainly based on two strict assumptions that practically cannot be satisfied in general. First, the asset returns are (jointly) normally distributed random variables. Second, return utility function is considered to be quadratic. It is widely accepted among researchers that, the rate of return in a portfolio is not jointly normal distributed.

A lot of research have improved Markowitz model in terms of computational and theoretical basis. Various measures of risk, such as half-variance model, the absolute deviation mean variance model, and the mean-variance-skewness model have been proposed that are discussed in the following sections. In this paper, the mean-variance-skewness model is used.

### D.2.1. Mean-Variance Model
According to Markowitz [3], the Mean- Variance model is formulated as:

$$\underset{X_{1:M}}{Min} \hat{V}(X_{1:M}) = \sigma_p^2 = \sum_{i=1}^{M} X_i^2 \sigma_i^2 + \sum_{i=1}^{M} \sum_{\substack{j=1 \\ j \neq i}}^{M} X_i X_j \operatorname{cov}(i,j)$$

(D.4)

$$S.t. \sum_{i=1}^{M} X_i R_i' = R_d'$$ 
(D.5)

$$\sum_{i=1}^{M} X_i = 1$$
(D.6)

$$X_i \geq 0, 1 \leq i \leq M$$
(D.7)

Where the cost function, $\hat{V}: R^M \to R$ incorporates the total risk of the portfolio, and $X_i, 1 \leq i \leq M$ are the weights of the stocks in the portfolio. Constraint (D.5) forces the portfolio return to have the desired return value. Constraint (D.6) shows the proper allocation of resources and the last constraint (D.7) indicates the sign of each weight. Therefore, mean-variance model poses a minimization problem that gives a set of optimal portfolios that has the minimum expected risk for a given expected return. Such an optimal set is called the efficient frontier, whose elements are called the efficient portfolios. Any efficient portfolio's specific characteristics is that, there is no lower-risk portfolio for the given return, while in its dual problem, there is no higher expected return portfolio for the given expected risk.

### D.3. The proposed model using Markowitz Model

The modern theories of portfolio selection have been derived from the Markowitz model, which shows the interrelationship between returns and risk. Since then, several portfolio selection models that consider returns and risks, such as the mean-variance model, have been developed. Most logical work has been done on the selection of portfolio by using the first two torques of distribution. Many researchers believe that only higher torques can be neglected if the distribution is symmetric (like Normal Distribution). Samuelson has shown that higher torques are important in portfolio selection for investors and almost all investors, in a selection of two portfolios, if the mean and variance are both equal, select the one that has a higher third torque. All of the above discussion convince us that skewness should be added to the mean-variance model as the third central torques. After researcher's studies [26], [27], because skewness is considered as the third torque in portfolio performance evaluation, in this paper, is tried to use the mean-variance-skewness model for the Iranian's stock market.

We build our portfolio optimization model based the Mean- Variance- Skewness model of Markowitz that considers variance as a measure of the risk. The original model is as follows:

$$Min_{w_{1:N}} \Rightarrow \lambda \sum_{i=1}^{N}\sum_{j=1}^{N} w_i w_j \sigma_{ij} - (1-\lambda)\sum_{i=1}^{N} w_i \mu_i - \theta \sum_{i=1}^{N} Skew(i) \quad (D.8)$$

$$\sum_{i=1}^{N} w_i = 1 \quad (D.9)$$

$$0 \leq w_i \leq 1, 1 \leq i \leq N \quad (D.10)$$

Where, $N$ is number of available assets, $w_i$ denotes the ratio of the $i$th asset in the portfolio, $\mu_i$ is the expected return on the $i$th asset, $\sigma_{ij}$ is the covariance of the $i$th and $j$th assets for $1 \leq i, j \leq N$, $\lambda$ and $\theta$ is weight coefficient for risk and Skewness and also $(1-\lambda)$ is weight coefficient for return, and $Skew(i)$ shows skewness of $i$th assets. The objective function (D.8) reflects the total objective (i.e., risk) of the portfolio, constraint (D.9) assures that the summation of the normalized weights is equal to 1, and (D.10) reflects the regularity condition on the ratio of each asset in the portfolio.

It is also very important in this model to obtain return and risk of portfolio which are shown in equations (D.11) and (D.12):

$$\mu_p = \sum_{i=1}^{N} w_i \mu_i \quad (D.11)$$

$$\sigma_p = \sum_{i=1}^{N}\sum_{j=1}^{N} w_i w_j \sigma_{ij} \quad (D.12)$$

### D.3.1. Expected Return and Risk of the Stocks

Let $R_t$ and $R_t'$ denote the real and predicted stock returns at time step $t$, respectively is obtained using the data of the last $t-1$ steps. Moreover, $\varepsilon_t = R_t - R_t'$ shows the prediction error at time step $t$, which is the difference between the real value and its prediction. Furthermore, the time series of the $N$ prediction errors is summarized as $\hat{\varepsilon} = (\varepsilon_1, \varepsilon_2, \varepsilon_3, ..., \varepsilon_N)$. In order to have a non-biased estimation, prediction errors must be statistically independent and identically distributed (i.i.d) with a mean and variance that are calculated as follows:

$$\mu_\varepsilon = \bar{\varepsilon} = 0 \quad (D.13)$$

$$\hat{V} = \sigma_\varepsilon^2 = \frac{1}{N-1} \sum_{t=1}^{t=N} \varepsilon_t^2 \quad (D.14)$$

Variance of the prediction error represents the uncertainty of the stock return. Therefore, it can be used as a criterion for measuring the risk of the stock. Generally, a higher variance indicates a higher risk.

### E. Taghochi Algorithm

In this section, we design a Taghochi algorithm for $\lambda=0.8, \theta=0.2$ in 5 indices portfolio optimization considering parameters like population size, selection function for initial population, crossover rate, crossover function and penalty factor on constraints that see the results below. Table (E.1) shows the amount of factors for different level of Taghochi Algorithm. Considering figure (E.1), the best estimate of population size is 200, roulette wheel is the best selection function for initial population, crossover rate must be 0.8 with single point crossover function and finally for penalty factor on constraints, the best penalty is 10. Using this experiment design, we use this for both 66 stocks and 5 indices per $\lambda$ and $\theta$.

Table E.1. The amount of factors for different level of Taghochi Algorithm

| Parameters | Level 1 | Level 2 | Level 3 |
|---|---|---|---|
| Population size | 50 | 100 | 200 |
| Selection function | Uniform | Roulette wheel | Tournament |
| Crossover fraction | 0.9 | 0.6 | 0.8 |
| Crossover function | Scattered | Single point | Two point |
| Penalty factor | 10 | 50 | 100 |

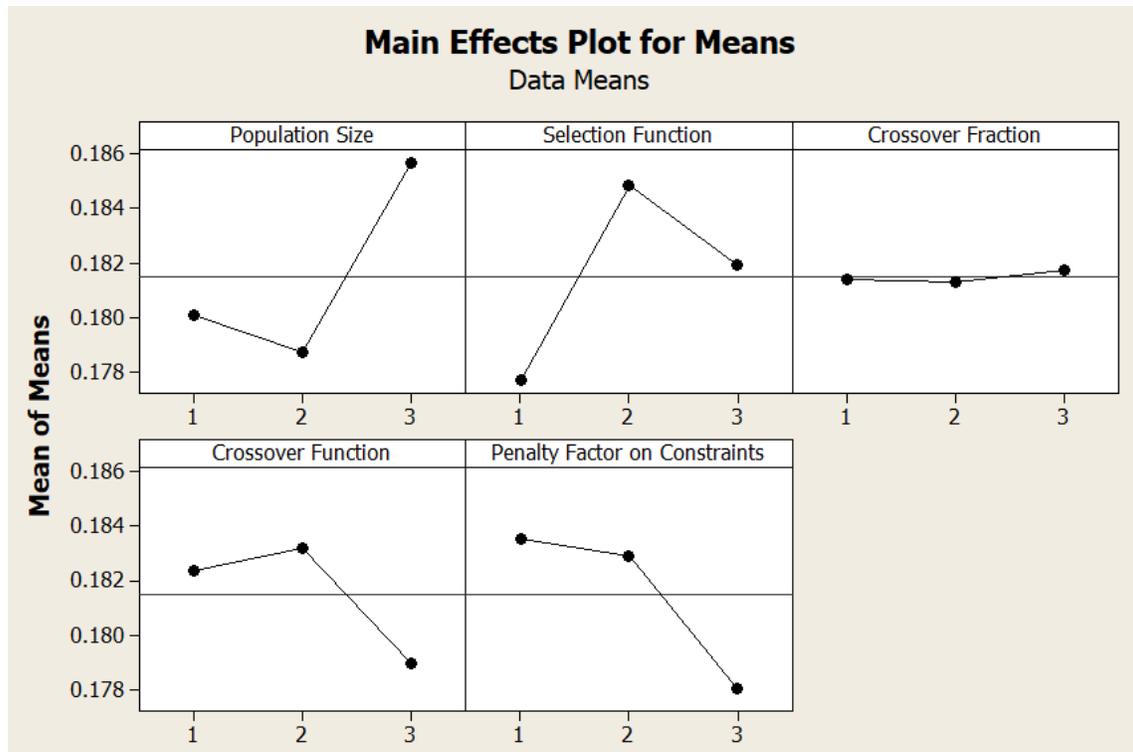

Figure E.1. The best estimate of parameters of genetic algorithm using Taghuchi algorithm

**F. Portfolio Optimization Using the Proposed Genetic Algorithm**
The genetic algorithm [45], is formed by an initial population of a fixed number of chromosomes that are randomly generalized. In the problems of portfolio optimization, each chromosome shows the weight of each stock in the portfolio and tries to lead to a feasible and optimal solution. In order to evaluate the fitness of each chromosome, an evaluation function is formed that shows how good each chromosome solution is. Using operations of crossover, mutation and natural selection (re-building of the population), the initial population converges towards solutions with better fitness. The better the fitness, the more optimal the value of the objective function. Let us provide the fundamental steps in the genetic algorithm. The first stage is to create a randomly generated population. The second step is to evaluate the fitness of each chromosome solution.

Next, using crossover, mutation and natural selection operators, genetic answers with low fitness are converged to developed chromosomes with higher fitness. Afterwards, the new population with higher fitness is replaced with the current population. In the fifth step, the stopping criteria of the algorithm is checked; if it is satisfied, the execution of the algorithm is terminated, otherwise, the algorithm is continued from the second step. In all of the rebuilding steps, the children of the two parent chromosomes are populated. Moreover, in all of the rebuilding cycles, the fittest solution becomes the best solution.

### F.1. Generation of the Population

In our problem, for higher efficiency of the solution, we generate a chromosome population of 200 members (using Taghochi Algorithm). Population type are selected double vector that required when there are integer constraints.

### F.2. Evaluating the Fitness of the Objective Function

We use the Mean-Variance-Skewness cost function:
$$f = \lambda \sum_{i \in Q} \sum_{j \in Q} w_i w_j \sigma_{ij} - (1-\lambda) \sum_{i \in Q} w_i \mu_i - \theta \sum_{i \in Q} Skew(i)$$
(E.1)

Presented chromosomes as a solution are composed of two distinct parts, a set of $Q$ number of $K$ types of stocks, and $K$ real numbers $s_i (0 \leq s_i \leq 1)$, for $1 \leq i \leq Q$. Share of all sets of $Q$ number of $K$ types of stock in portfolio is $\sum_{i \in Q} \varepsilon_i$. $s_i$ Is the free portion of the portfolio which is calculated as $1 - \sum_{i \in Q} \varepsilon_i$. Therefore, the weight allocated to the $i$th asset in portfolio is:

$$w_i = \varepsilon_i + (\frac{s_i}{\sum_{i \in Q} s_i})(1 - \sum_{i \in Q} \varepsilon_i)$$
(E.2)

Note that only some of the generated chromosomes are feasible due to the constraints on equation (D.10). In evaluation of the genetic algorithm all restrictions on the lower limit of $\varepsilon_i$ are automatically established. However, in order to impose the upper limits of $\delta_i$, a creative method is needed. [Lower limit of optimal portfolio of 5 indices =0.1 vs upper limit of optimal portfolio of 5 indices =0.3, lower limit of optimal portfolio of 66 stocks =0.1 vs upper limit of optimal portfolio of 66 stocks =0.3]

## F.3. Operations of Crossover, Mutation and Natural selection

In this subsection, for the natural selection has been used Roulette wheel method (using Taghochi Algorithm). This method uses a random number to select one of the sections with a probability equal to this area. We describe how the genetic operators in the algorithm are formed and modified. In this paper, children in the genetic algorithm are generated using two-point crossover operator. Using two-point crossover operator two parents only produce one child. The algorithm selects genes numbered less than or equal to $m$ from the first parent, selects genes numbered from $m+1$ to $n$ from the second parent, and select genes numbered greater than $n$ from the first parent. The algorithm then concentrates these genes to from a single gene. If the stock $i$ in both parents are alike with the same features, those features will be shown on child with allocation value of $s_i$ which is chosen randomly from either of the parents. If the $i$th stock is only present in one of the parents, it is present with probability 0.5 in the child as well. Also crossover fraction is 0.9 in this paper. Moreover, in the mutation operator, are used adaptive feasible that randomly generates directions that are adaptive with respect to the last successful or unsuccessful generation. A step length is chosen along each direction so that linear constraints and bounds are satisfied.

## F.4. Constraints Penalty

In this subsection, we are used of initial penalty and penalty factor for constraints. For initial penalty have established 10 that specifies an initial value to be used by the algorithm. Also penalty factor increases the penalty parameter when the problem is not solved to required accuracy and constraints are not satisfied. In this article Penalty factor has been 10(Using Taghuchi Algorithm).

## F.5. Replacement

The algorithm uses a fixed population replacement strategy. In this strategy, when a child is generated, it is replaced in the population according to the value of the objective function.

## F.6. Stopping Criteria

We use *1000* second running time for the heuristic genetic algorithm in a fixed $\lambda$ and $\theta$. The average value of all the repetitions are used as the value for each $\lambda$ and $\theta$. Moreover, if the weighted average change in the fitness function value over stall generations is less than function tolerance, the algorithm stops. Stall generation and function tolerance specified 50 and $1\times10^{-6}$.

## F.7. Flowchart

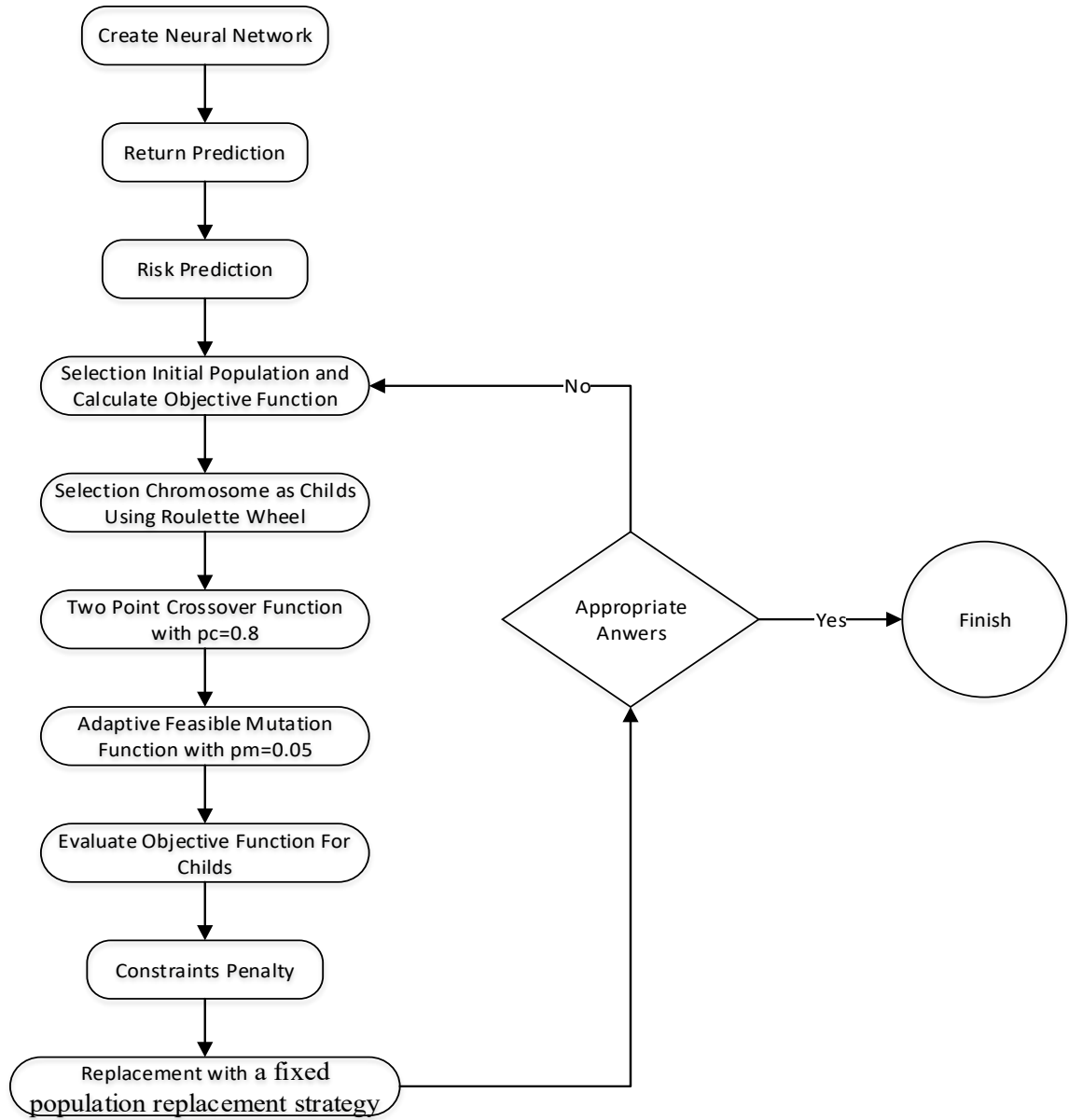

## G. Model Analysis and Results

In this section, we perform some tests to evaluate the neural network predictors and to show that the prediction errors have normal distributions. Finally, we will perform tests to show the performance of the prediction based portfolio optimization model on real data obtained under different risks. Moreover, in this section we determine the return and risk of the set of 66 stocks in the five indices using neural network.
Then, we obtain the return, risk and correlation of five indices and 66 stocks using neural network. Lastly, using meta-heuristic optimization approach based on genetic algorithms, weight of each indices and stocks specified, plotted analyzed for 4 types of $\lambda$ and 2 types of $\theta$.

### G.1. Data

In this study, from 21 indices in Iran's stock market, we select five leading indices including bank, insurance, investment, oil and chemicals. From different stocks in each of these selected indices, some stocks with few historical data have been removed. So 66 stocks in these indices are selected for evaluation and creation of an active portfolio. For each of the 66 stocks, we obtained 221 weekly returns between the time period beginning 03/01/2012 and the 07/16/2016 with closed prices on Monday (to eliminate the effect of first and last day of the week). In some stocks, there were also missing data that in all cases the final price of the day before was used. Table (F.1) represents the used stocks in each index.

Table G.1. All of the stocks of the 5 indices used in the experiment

| Bank | Insurance | Investment | Petroleum Materials | Chemical Materials |
|---|---|---|---|---|
| Ansar | Alborz | Gostaresh | Bahman | Alyaf Masnui |
| Day | Asia | Mellat | Pars | Carbon Iran |
| Eghtesad Novin | Dena | Pardis | Palayesh Naft Tabriz | Goltash |
| Gardeshgari | Day | Atieh Damavand | Palayesh Naft Esfahan | Lubiran |
| Hekmat Iranian | Parsian | Bahman | Sanat Naft | Marun |
| Karafarin | Pasargad | Bu-Ali | | Rang Niru |
| Khavarmianeh | Saman | Gruh Behshahr | | Paksan |
| Mellat | Mellat | Kharazmi | | Parsan |
| Parsian | Mihan | Melli | | Petrol |
| Saderat | Veskab | Saipa | | Petroshimi Abadan |
| Sarmayeh | Takado | Sanat Nimeh | | Petroshimi Fanavaran |
| Sina | | Sanat & Madan | | Petroshimi Kermanshah |
| Tejarat | | Sepah | | Petroshimi Khark |
| Pasargad | | Tose-E Melli | | Petroshimi Khark |
| PostBankIran | | Tose-E Sanat | | Petroshimi Pardis |
| | | | | Petroshimi Shazand |
| | | | | Petroshimi Shiraz |
| | | | | Poly Cril Iran |
| | | | | Sanaye Shimyayi Iran |
| | | | | Shimyayi Fars |

### G.2. Return Prediction

The training parameters of the predictive neural network in the previous works were evaluated experimentally [46]. In this paper, a nonlinear autoregressive (NAR) [47] with a feedforward neural network containing a hidden layer with the learning function of Levenberg-Marquardt (trainlm) with a topology of 1:5:1 (1 input neuron, 5 neurons in hidden layer and 1 output neuron) is used. Figure (G.1) and Figure (G.2) show topology of the nonlinear autoregressive (NAR) and Neural Network that have been used for prediction of the returns. As it is seen from Figure (F.1), this predictive, takes past *d* data(returns) as input to predict the future data (future return) by using a feedback. In other words, it counts this *d* data as delay in neural network prediction.

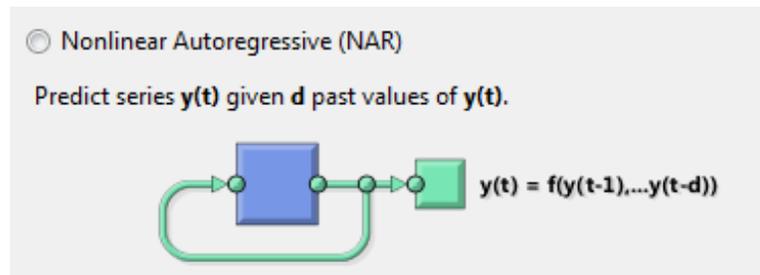

Figure G.1. Topology of the nonlinear predictor.

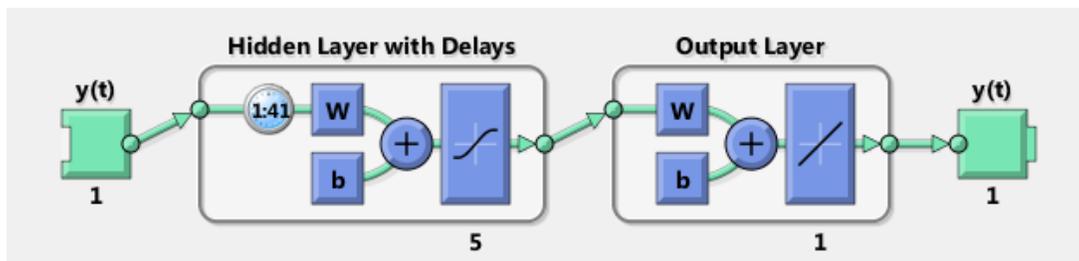

Figure G.2. Topology of the used neural network.

For training and testing the neural networks in the experiments, of the 221 weekly returns, 41 returns of the first weekly returns has been deleted as delays in neural network. So, we will have 11,880 sessions on neural network (180 learning cycle and testing 66 stocks, 11880=180*66). Each learning session in the neural network has been done on 1000 epic with the Levenberg-Marquardt learning algorithms. Of the 221 returns, %70 have been used as training, %15 as validation and the remaining %15 as testing. So learning section will have 155 data and each of the validation and testing sections will have 33 data. To reduce the over fitting, we have used the known technique in neural network that has two sections, one section for training and another section for validation. Training section is used to update the weights of the neural network while the validation section tries to pick up in 1000 epic of training. The traditional use of the training and validation for controlling over fitting has been challenged by moving time series prediction [48-52]. Although, there is no standard method for the validation section but there is some of methods such as transfer to other sections. In this paper, we used heuristic methods in the work of HAYKIN for the training and validation sections [48].

### G.3. Evaluation Standards

We have used mean error (ME), root mean square error (RMSE), mean absolute percentage error (MAPE) and Standards of hit rate to evaluate the performance of our prediction [53]. Mean error is the difference between real return and the predicted return and is calculated as follows:

$$ME = \frac{1}{n}\sum_{i=1}^{n}\left|R_t - \hat{R}_t\right|$$

(G.1)

Where $n$ is the length of time series, $R_t$ and $\hat{R}_t$ are real and predicted returns in time $t$. Mean error is used to evaluate the assumption $\mu_\varepsilon = \bar{\varepsilon} = 0$ and normality of the prediction error.

Root mean square error is an evaluation standard to compare differences between two-time series and is defined as follows:

$$RMSE = \sqrt{\frac{1}{n}\sum_{i=1}^{n}(R_t - \hat{R}_t)^2}$$

(G.2)

Root mean square error is interpreted as the standard deviation of prediction errors. Given the mean of zero and the second root of the mean error, the distance between these errors and the ideal state of the mean zero error is shown. RMSE is sensitive for small deviations in the data. Mean absolute percentage error is defined as follows:

$$MAPE = \frac{1}{n}\sum_{i=1}^{n}\frac{\left|R_t - \hat{R}_t\right|}{R_t}$$

(G.3)

The average percentage error is also a free measure of the unit and for $R_t$ is very close to zero and $\hat{R}_t$ the opposite of its, *MAPE* can be found to be the average of large errors.

Hit rates are defined as follows and measure the percentage of the predictions that signal $R_t$ and $\hat{R}_t$ are consistent with each.

$$H_R = \frac{Count_{t=1}^{n}(R_t \cdot \hat{R}_t > 0)}{Count_{t=1}^{n}(R_t \cdot \hat{R}_t \neq 0)}$$

(G.4)

$$H_{R^+} = \frac{Count_{t=1}^{n}(R_t > 0 \, AND \, \hat{R}_t > 0)}{Count_{t=1}^{n}(R_t > 0)}$$

(G.5)

$$H_{R^-} = \frac{Count_{t=1}^{n}(R_t < 0 \, AND \, \hat{R}_t < 0)}{Count_{t=1}^{n}(R_t < 0)}$$

(G.6)

$H_R$ represents the percentage of times that the signals $R_t$ and $\hat{R}_t$ are non-zero and have the same signal (both of them + or both -). $H_{R^+}$ and $H_{R^-}$ represents the percentage of the predictions of the signals $R_t$ and $\hat{R}_t$ that both are positive and negative respectively.

### G.4. Prediction Performance

In table (G.2), we present the experimental results and evaluate the prediction performance for all of the 66 stocks with neural networks. This table shows the mean, variance and standard deviation for the results of the 221 prediction of each predictor in accordance with the standards discussed previously. In this table, we see that ME is so close to zero with very low standard deviation. $H_{R^+}$ has the performance of %64 which is %14 better than the random selection of positive return of the market. The portfolio optimization model is based on our prediction, which will be shown in the next section, has a good portfolio of functions.

Table G.2. A summary of the results obtained for 66 stocks and the evaluation of defined standards

| Error | Mean | $\sigma^2$(Variance) | $\sigma$(Standard Deviation) |
|---|---|---|---|
| ME | 0.000432 | 0.000031 | 0.005567 |
| RMSM | 0.040902 | 0.000543 | 0.0233023 |
| MAPE | 6.397514 | 120.832151 | 10.992367 |
| $H_R$ | 0.536423 | 0.007632 | 0.087361 |
| $H_{R^+}$ | 0.647894 | 0.009183 | 0.095827 |
| $H_{R^-}$ | 0.36562 | 0.009323 | 0.096555 |

### G.5. Normality of Prediction errors and Returns

We have used the normality of prediction errors and returns of accurate time series for portfolio optimization. We tested a percentage of these series by Kolmogorov- Smirnov test and accepted the normality (otherwise rejected). For the normality test, all the 66 stocks in the portfolio were tested between 16[th] of August, 2014 till 16th of July, 2016 for 100 weeks. The results of the normality tests for returns and prediction errors are represented in figure (G.3) and (G.4) respectively. Normality index of prediction errors is less variable than normality index for series of returns. Figure (G.5) is a comparison between time series of returns and prediction errors.

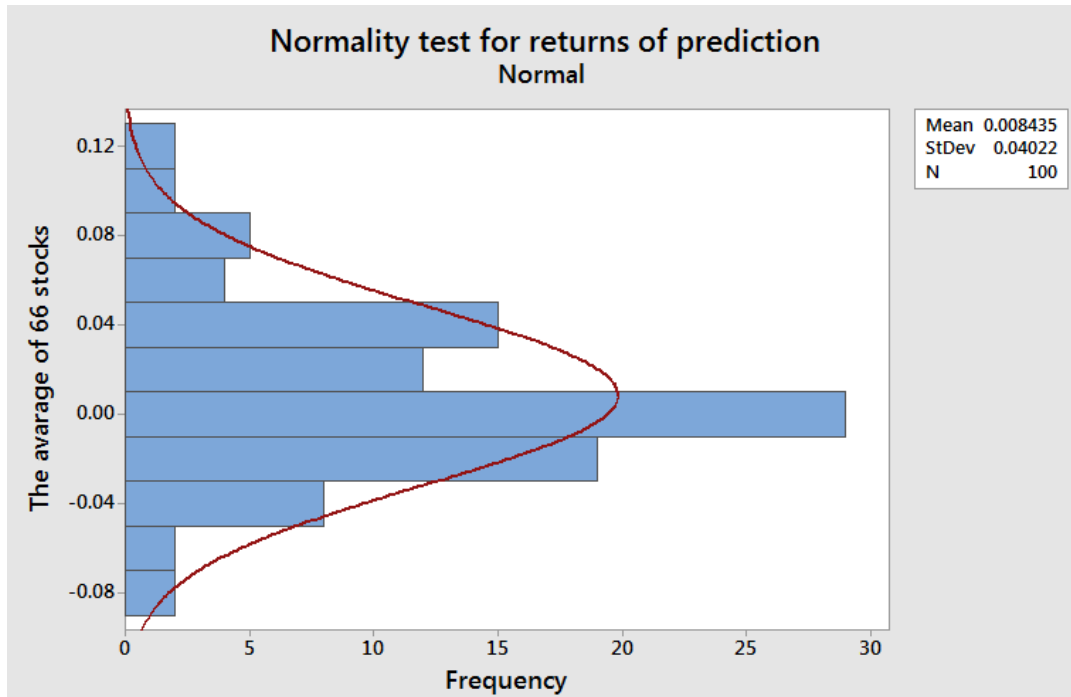

Figure G.3. Normality test for predicted returns of time series.

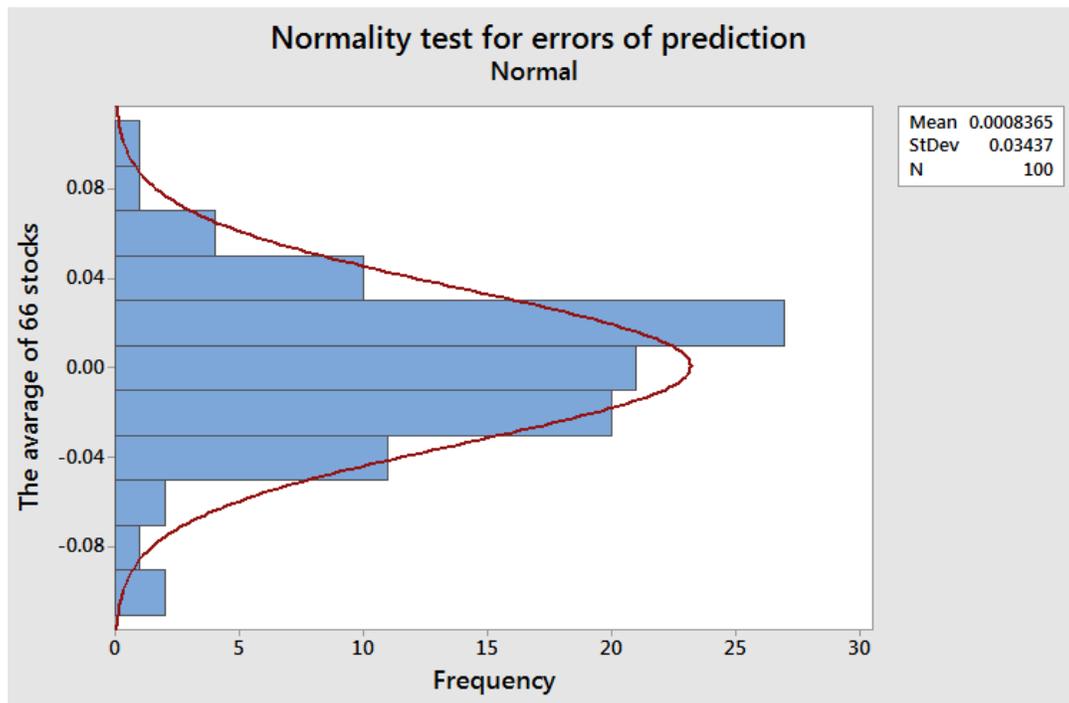

Figure G.4. Normality test for predicted error series

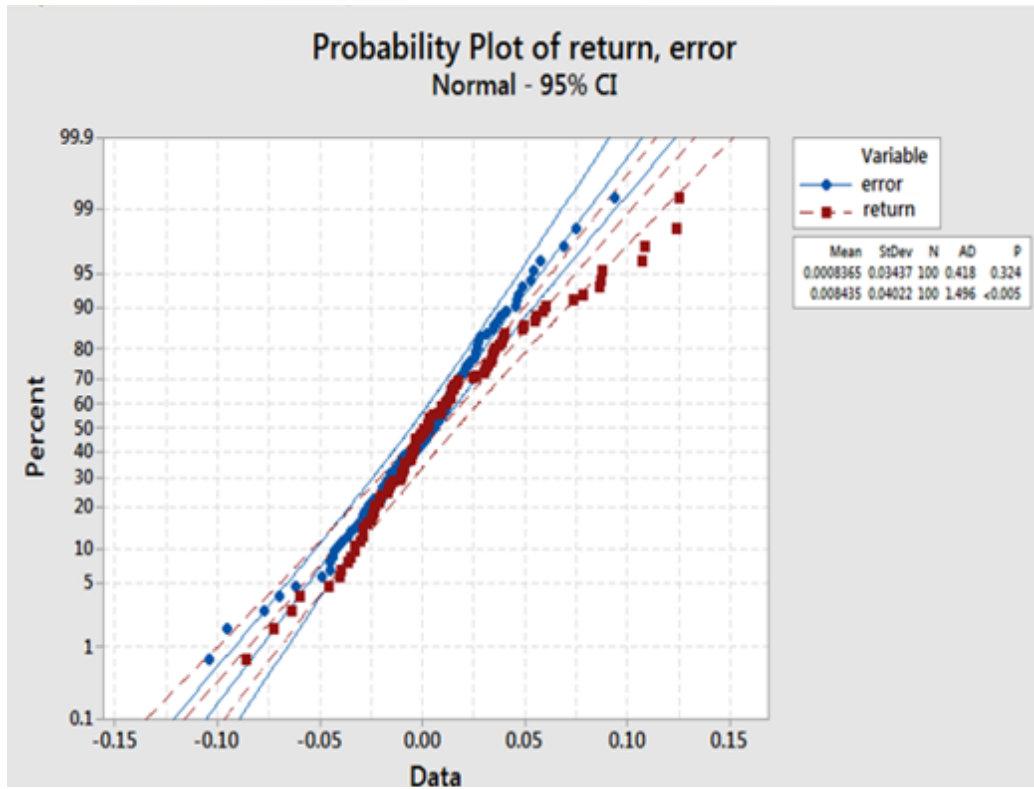

Figure G.5. Normality test for predicted errors and returns for %95 confidence level.

Table (G.3) briefly describes 100 normality index for each time series with mean, variance and standard deviation of normality index.

Table G.3. The summary of normality index for predicted returns and predicted error series

|  | Mean | $\sigma^2$ (Variance) | $\sigma$ (Standard Deviation) |
|---|---|---|---|
| **normality index for predicted returns** | 0.000836 | 0.001181 | 0.03437 |
| **normality index for predicted error series** | 0.008435 | 0.001617 | 0.04022 |

As can be seen, we can not only use the mean and the variance for stocks in the Iran's stock market and skewness should also be used as a third element. So, we cannot consider the returns and the series of predictive errors as normal.

### G.6. Predicting stocks risk and return using neural network

In this section we use neural network for each of the stocks of the five indices containing bank, insurance, investment, petroleum Materials and chemical Materials in the time periods of 180 weeks to specify return and risk of each stock. (Table G.4- G-9)

Table G.4. Rerun and risk stocks in 5 Indices

| 5 Indices | Return | Risk/ $\sigma^2$(Variance) |
|---|---|---|
| **Bank** | 0.004916 | 0.002834 |
| **Insurance** | 0.005032 | 0.003145 |
| **Investment** | 0.009823 | 0.005467 |
| **Petroleum Materials** | 0.009700 | 0.003454 |
| **Chemical Materials** | 0.003187 | 0.003478 |

Table G.5. Rerun and risk stocks in Bank Index

| **Bank** | **Return** | **Risk/ $\sigma^2$(Variance)** |
|---|---|---|
| Ansar | 0.007706 | 0.001110 |
| Day | 0.003314 | 0.001961 |
| Eghtesad Novin | 0.002809 | 0.001483 |
| Gardeshgari | 0.013410 | 0.004332 |
| Hekmat Iranian | 0.012764 | 0.002312 |
| Karafarin | 0.002811 | 0.000993 |
| Khavarmianeh | 0.019691 | 0.002081 |
| Mellat | 0.006980 | 0.001599 |
| Parsian | 0.004730 | 0.001315 |
| Saderat | 0.007848 | 0.001871 |
| Sarmayeh | 0.001122 | 0.004208 |
| Sina | 0.003972 | 0.001466 |
| Tejarat | 0.001816 | 0.001629 |
| Pasargad | 0.006176 | 0.000937 |
| PostBankIran | 0.009452 | 0.003034 |

Table G.6. Rerun and risk stocks in Insurance Index

| **Insurance** | **Return** | **Risk/ $\sigma^2$(Variance)** |
|---|---|---|
| Alborz | 0.007459 | 0.001931 |
| Asia | -0.000380 | 0.001932 |
| Dena | 0.003145 | 0.003879 |
| Day | -0.000780 | 0.001170 |
| Parsian | 0.011215 | 0.001036 |
| Pasargad | 0.009887 | 0.000607 |
| Saman | 0.011553 | 0.003089 |
| Mellat | 0.003482 | 0.001769 |
| Mihan | 0.002007 | 0.003368 |
| Veskab | 0.005182 | 0.003002 |
| Takado | 0.007045 | 0.004237 |

Table G.7. Rerun and risk stocks in Investment Index

| Investment | Return | Risk/ $\sigma^2$(Variance) |
|---|---|---|
| Gostaresh | 0.012226 | 0.005212 |
| Mellat | -0.003090 | 0.004024 |
| Pardis | -0.000670 | 0.001543 |
| Atieh Damavand | 0.001576 | 0.001243 |
| Bahman | 0.013591 | 0.001038 |
| Bu-Ali | 0.010363 | 0.002977 |
| Gruh Behshahr | -0.002110 | 0.001417 |
| Kharazmi | -0.010180 | 0.001514 |
| Melli | 0.002546 | 0.001780 |
| Saipa | -0.008190 | 0.002632 |
| Sanat Nimeh | 0.007237 | 0.002642 |
| Sanat & Madan | 0.014082 | 0.001432 |
| Sepah | 0.010517 | 0.001389 |
| Tose-E Melli | 0.008655 | 0.000915 |
| Tose-E Sanat | -4.99E-03 | 0.002093 |

Table G.8. Rerun and risk stocks in Petroleum Materials Index

| Petroleum Materials | Return | Risk/ $\sigma^2$(Variance) |
|---|---|---|
| Bahman | 0.010733 | 0.001767 |
| Pars | 0.015436 | 0.003062 |
| Palayesh Naft Tabriz | 0.008882 | 0.003067 |
| Palayesh Naft Esfahan | 0.001479 | 0.001917 |
| Sanat Naft | 0.012399 | 0.001901 |

Table G.9. Rerun and risk stocks in Chemicals Materials Index

| Chemical Materials | Return | Risk/ $\sigma^2$(Variance) |
|---|---|---|
| Alyaf Masnui | -0.003640 | 0.005458 |
| Carbon Iran | -0.012660 | 0.004366 |
| Goltash | 0.022769 | 0.003057 |
| Lubiran | -8.86655E-05 | 0.002304 |
| Marun | 0.007800 | 0.001765 |
| Rang Niru | -0.002800 | 0.001219 |
| Paksan | 0.014008 | 0.001306 |
| Parsan | 0.009011 | 0.001250 |
| Petrol | 0.010418 | 0.003478 |
| Petroshimi Abadan | 0.003361 | 0.001036 |
| Petroshimi Fanavaran | 0.011858 | 1.38E-03 |
| Petroshimi Kermanshah | 0.013970 | 0.001106 |
| Petroshimi Khark | 0.007589 | 0.001543 |
| Petroshimi Khark | 0.005022 | 0.001323 |
| Petroshimi Pardis | 0.018282 | 0.001709 |
| Petroshimi Shazand | 0.008511 | 0.001534 |

| | | | | | | | | | | |
|---|---|---|---|---|---|---|---|---|---|---|
| Petroshimi Shiraz | | | -0.010270 | | | | | 0.003967 | | |
| Poly Cril Iran | | | 0.010647 | | | | | 0.002693 | | |
| Sanaye Shimyayi Iran | | | 0.005551 | | | | | 0.003241 | | |
| Shimyayi Fars | | | 0.024780 | | | | | 0.004461 | | |

### G.7. Determining investment amounts considering minimum variance ($\lambda=1$), maximum return ($\lambda=0$), high risk people ($\lambda=0.2$) and risk aversion people ($\lambda=0.8$) of the 5 indices and of the 66 stocks

In this section, after obtaining the values of return, variance, covariance and skewness of each of the 5 indices and 66 stocks in these indices, which are designed using the neural network. For amount of $\lambda=1, 0.8, 0.2, 0$ and $\theta=0, 0.2, 0.8$ that we get an optimal portfolio for each of the sections that the results are shown in the table (G.10) and (G.11)

Table G.10. Weight coefficient for 5 indices for optimal portfolio with different $\lambda$ and $\theta$

| Weight coefficients for 5 Indices | $\lambda=1$ $\theta=0$ | $\lambda=1$ $\theta=0.2$ | $\lambda=1$ $\theta=0.8$ | $\lambda=0.8$ $\theta=0.2$ | $\lambda=0.8$ $\theta=0.8$ | $\lambda=0.2$ $\theta=0.2$ | $\lambda=0.2$ $\theta=0.8$ | $\lambda=0$ $\theta=0$ | $\lambda=0$ $\theta=0.2$ | $\lambda=0$ $\theta=0.8$ |
|---|---|---|---|---|---|---|---|---|---|---|
| Bank | 0.131 | 0.298 | 0.3 | 0.299 | 0.298 | 0.299 | 0.3 | 0.16 | 0.105 | 0.102 |
| Insurance | 0.219 | 0.145 | 0.16 | 0.241 | 0.26 | 0.156 | 0.256 | 0.1 | 0.090 | 0.66 |
| Investment | 0.089 | 0.1 | 0.1 | 0.1 | 0.1 | 0.1 | 0.1 | 0.38 | 0.298 | 0.380 |
| Petroleum Materials | 0.069 | 0.286 | 0.284 | 0.204 | 0.177 | 0.29 | 0.179 | 0.32 | 0.423 | 0.389 |
| Chemical Materials | 0492 | 0.172 | 0.157 | 0.156 | 0.165 | 0.156 | 0.166 | 0.04 | 0.084 | 0.063 |
| Rerun of Portfolio ($\mu_p$) | 0.0049 | 0.0065 | 0.0065 | 0.0061 | 0.0060 | 0.0065 | 0.0060 | 0.0082 | 0.0083 | 0.0085 |
| Risk of Portfolio ($\sigma_p$) | 0.0005 | 0.0008 | 0.0008 | 0.0007 | 0.0006 | 0.0009 | 0.0006 | 0.0012 | 0.0014 | 0.0015 |

Figure (G.6) shows the efficient frontier between the five indicators ($\theta=0$, that's mean regardless of the effect of skewness in the objective function, efficient frontier is specified) which according to the data in table (G.10), the highest return of this portfolio is 0.008190 ($\theta=0, \lambda=0$, that's mean regardless of the effect of skewness and variance in the objective function, and that the most important problem is to increase the desired return) and also the return related to minimum variance of this portfolio is 0.004878 ($\theta=0, \lambda=1$, that's mean regardless of the effect of skewness and return in the objective function, and that the most important problem is to decrease the desired risk or desired variance). In other words, when the investor decides to select a portfolio of these five indicators, he should know that achieving returns of more than 0.008190 per week and achieving a risk (or variance) when the return is less than 0.004878 per week is not possible. In addition to the results, according to table (G.10), in each of the different values of $\lambda$, the $\theta$ increases (That's mean, the effect of skewness in the objective function is greater), the return and risk of the desired portfolio are also increased.

Consequently, skewness can be considered as an inseparable element of the model. Now, according to risk lover investors or risk aversion investors, different returns can be achieved with different risks. Perhaps some investors are looking for high returns with high risk, and some are looking for at low return with low risk. It is clear that the investors just do not like the highest returns and least variance, and they are looking for more diversity. According to figure (G.6), different portfolios can be identified that have different risk and return and the weight of each index isn't equal in different portfolios.

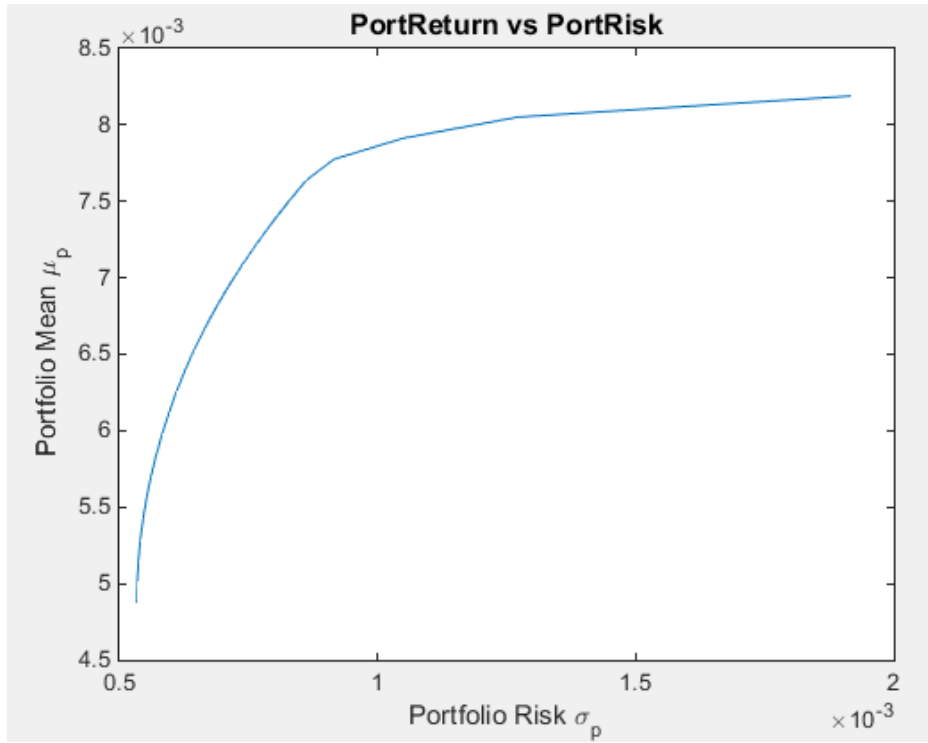

Figure G.6. Efficient Frontier of 5 Indices

Also figure (G.7) shows the efficient frontier between the 66 stocks that which according to the data in table (G.11), the highest return of this portfolio is 0.027794 ($\theta = 0, \lambda = 0$) and also the return related to minimum variance of this portfolio is 0.008416 ($\theta = 0, \lambda = 1$). In other words, when the investor decides to select a portfolio of these five indicators, he should know that achieving returns of more than 0.027794 per week and achieving a risk (or variance) when the return is less than 0.008416 per week is not possible.

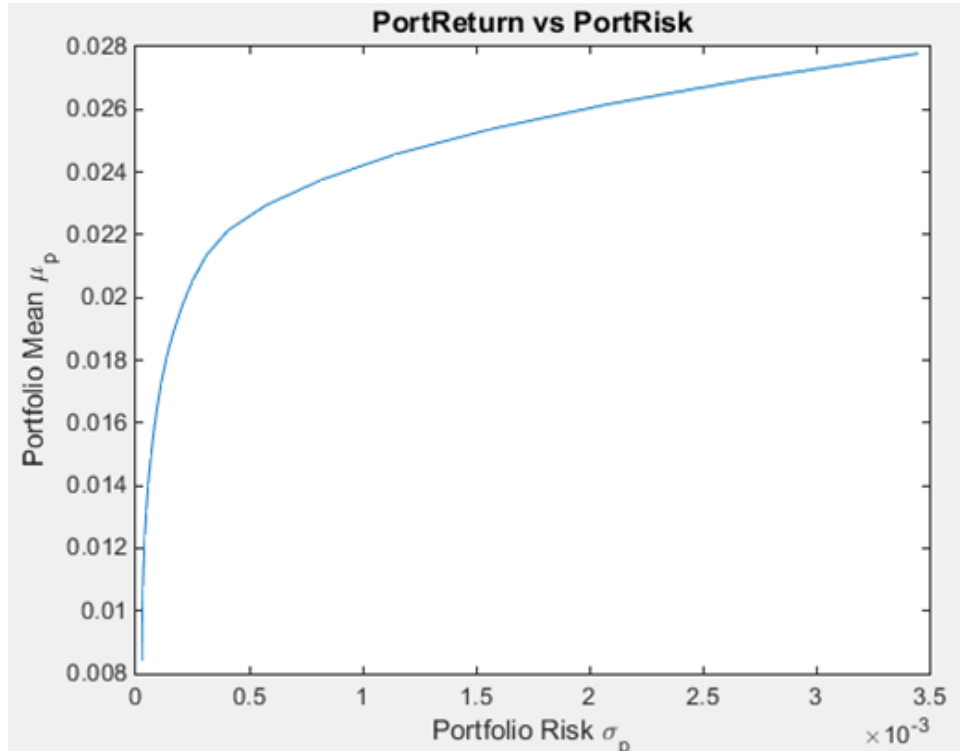

Figure G.7. Efficient Frontier of 66 stocks

**H. Conclusion and Future Work**
In this paper, after defining the overall problem and weakness of investment in Iran, a mathematical model was proposed to obtain risk and return of the market by using Neural Network. There was some experiment to test the normality of predicted return and errors and by using evaluation standards, we saw that our model is so close to the Mean-Variance-Skewness with weight constraints Model. Finally, the proposed model was implemented on real data of Iran's stock market and it was concluded that buying single stock is better than buying an index. As future work, we will implement the proposed method for all the 21 indices of Iran's market. Objective function in this work was Mean-Variance. As future work we could use Mean-Variance-Skewness-kurtosis with liquidity constraints or use the methods that measure risks like value in the risk.

## K. Key References

[1] Fabio D. Freitas, Alberto F. De Souza, Alison R. De Almeida. Prediction-based portfolio optimization model using neural networks, Neuro computing 72 (2009) 2155–2170.

[2] Ali Husseinzadeh Kashan. League Championship Algorithm (LCA): An algorithm for global optimization inspired by sport championships. Applied Soft Computing 16 (2014) 171–200.

## L. References

[1] Edwin J Elton, Martin J Gruber, SJ Brown, and WN Goetz Mann. Modern portfolio theory and investment analysis, john iley&sons. Inc., New York, 1995.

[2] Harry Markowitz. Portfolio selection. The journal of finance, 7(1):77– 91, 1952.

[3] Konno, H., & Suzuki, K. I. (1995). A mean-variance-skewness portfolio optimization model. Journal of the Operations Research Society of Japan, 38(2), 173-187.

[4] Yoshimoto, Atsushi. "The mean-variance approach to portfolio optimization subject to transaction costs." Journal of the Operations Research Society of Japan 39, no. 1 (1996): 99-117.

[5] Soleimani, Hamed, Hamid Reza Golmakani, and Mohammad Hossein Salimi. "Markowitz-based portfolio selection with minimum transaction lots, cardinality constraints and regarding sector capitalization using genetic algorithm." Expert Systems with Applications 36, no. 3 (2009): 5058-5063.

[6] Fernández, Alberto, and Sergio Gómez. "Portfolio selection using neural networks." Computers & Operations Research 34, no. 4 (2007): 1177-1191.

[7] Konno, Hiroshi, and Hiroaki Yamazaki. "Mean-absolute deviation portfolio optimization model and its applications to Tokyo stock market." Management science 37, no. 5 (1991): 519-531.

[8] Vercher, Enriqueta, and Jose D. Bermudez. "A possibilistic mean-downside risk-skewness model for efficient portfolio selection." IEEE Transactions on Fuzzy Systems 21, no. 3 (2013): 585-595.

[9] Harry M Markowitz. Portfolio selection: efficient diversification of investments, volume 16. Yale university press, 1968.

[10] John Y Campbell, Andrew Wen-Chuan Lo, Archie Craig MacKinlay, et al. The econometrics of financial markets, volume 2. Princeton University press Princeton, NJ, 1997.

[11] Daniel Bienstock. Computational study of a family of mixed-integer quadratic programming problems. In Integer programming and combinatorial optimization, pages 80–94. Springer, 1995.

[12] T-J Chang, Nigel Meade, John E Beasley, and Yazid M Sharaiha. Heuristics for cardinality constrained portfolio optimization. Computers & Operations Research, 27(13):1271–1302, 2000.

[13] Thomas Goll and Jan Kallsen. Optimal portfolios for logarithmic utility. Stochastic processes and their applications, 89(1):31–48, 2000.

[14] Teresa León, Vicente Liern, and Enriqueta Vercher. Viability of infeasible portfolio selection problems: A fuzzy approach. European Journal of Operational Research, 139(1):178–189, 2002.

[15] Yves Crama and Michael Schyns. Simulated annealing for complex portfolio selection problems. European Journal of operational research, 150 (3):546–571, 2003.

[16] E Ammar and HA Khalifa. Fuzzy portfolio optimization a quadratic programming approach. Chaos, Solitons & Fractals, 18(5):1045–1054, 2003.

[17] Xiao Lou Yang. Improving portfolio efficiency: A genetic algorithm approach. Computational Economics, 28(1):1–14, 2006.

[18] Mingming Liu and Yan Gao. An algorithm for portfolio selection in a frictional market. Applied mathematics and computation, 182(2):1629– 1638, 2006.

[19] Guang-Feng Deng, Woo-Tsong Lin, and Chih-Chung Lo. Markowitz based portfolio selection with cardinality constraints using improved particle swarm optimization. Expert Systems with Applications, 39(4):4558– 4566, 2012.

[20] Xiaohua Ma, Yuelin Gao, and Bo Wang. Portfolio optimization with cardinality constraints based on hybrid differential evolution. AASRI Procedia, 1:311–317, 2012.

[21] Rosella Castellano, Roy Cerqueti. Mean–Variance portfolio selection in presence of infrequently traded stocks. European Journal of Operational Research, 234(2), 442-449, 2014.

[22] Edwin J Elton, Martin J Gruber, Stephen J Brown, and William N Goetz Mann. Modern portfolio theory and investment analysis. John Wiley & Sons, 2009.

[23] Eugene F Fama. Portfolio analysis in a stable partial market. Management science, 11(3):404–419, 1965.

[24] William F Sharpe, Gordon J Alexander, and Jeffery V Bailey. Investments, volume 6. Prentice-Hall Upper Saddle River, NJ, 1999.

[25] Burton G Malkiel. The efficient market hypothesis and its critics. Journal of economic perspectives, pages 59–82, 2003.

[26] Alex Edmans, Diego Garcia, and Øyvind Norli. Sports sentiment and stock returns. The Journal of Finance, 62(4):1967–1998, 2007.

[27] Alberto Fernández and Sergio Gómez. Portfolio selection using neural networks. Computers & Operations Research, 34(4):1177–1191, 2007.

[28] Tunchan Cura. Particle swarm optimization approach to portfolio optimization. Nonlinear Analysis: Real World Applications, 10(4):2396– 2406, 2009.

[29] Lean Yu, Shou yang Wang, and Kin Keung Lai. Neural network-based mean–variance–skewness model for portfolio selection. Computers & Operations Research, 35(1):34–46, 2008.

[30] M Grazia Speranza. A heuristic algorithm for a portfolio optimization model applied to the Milan stock market. Computers & Operations Research, 23(5):433–441, 1996.

[31] Xiaoxia Huang. Mean-semi variance models for fuzzy portfolio selection. Journal of computational and applied mathematics, 217(1):1–8, 2008.

[32] Wei Yan, Rong Miao, and Shurong Li. Multi-period semi-variance portfolio selection: Model and numerical solution. Applied Mathematics and Computation, 194(1):128–134, 2007.

[33] M Arenas Parra, A Bilbao Terol, and MV Rodrıguez Urıa. A fuzzy goal programming approach to portfolio selection. European Journal of Operational Research, 133(2):287–297, 2001.

[34] Ismail Erol and William G Ferrell. A methodology for selection problems with multiple, conflicting objectives and both qualitative and quantitative criteria. International Journal of Production Economics, 86(3):187–199, 2003.

[35] OLV Costa and AC Paiva. Robust portfolio selection using linear-matrix inequalities. Journal of Economic Dynamics and Control, 26(6):889–909, 2002.

[36] Masaaki Ida. Portfolio selection problem with interval coefficients. Applied Mathematics Letters, 16(5):709–713, 2003.

[37] Matthias Ehrgott, Kathrin Klamroth, and Christian Schwehm. An MCDM approach to portfolio optimization. European Journal of Operational Research, 155(3):752–770, 2004.

[38] Chorng-Shyong Ong, Jih-Jeng Huang, and Gwo-Hshiung Tzeng. A novel hybrid model for portfolio selection. Applied Mathematics and computation, 169(2):1195–1210, 2005.

[39] Xiao-Tie Deng, Zhong-Fei Li, and Shou-Yang Wang. A minimax portfolio selection strategy with equilibrium. European Journal of operational research, 166(1):278–292, 2005.

[40] Chi-Ming Lin and Mitsuo Gen. An effective decision-based genetic algorithm approach to multi objective portfolio optimization problem. Applied Mathematical Sciences, 1(5):201–210, 2007.

[41] KP Anagnostopoulos and G Mamanis. A portfolio optimization model with three objectives and discrete variables. Computers & Operations Research, 37(7):1285–1297, 2010.

[42] Taras Bodnar, Nestor Parolya, and Wolfgang Schmid. On the equivalence of quadratic optimization problems commonly used in portfolio theory. European Journal of Operational Research, 229(3):637–644, 2013.

[43] Thomas HellstrSom and Kenneth HolmstrSom. Predicting the stock market. 1998.

[44] Eugene F Fama. Efficient capital markets: A review of theory and empirical work. The journal of Finance, 25(2):383–417, 1970.

[45] Darrell Whitley. A genetic algorithm tutorial. Statistics and computing, 4(2):65–85, 1994.

[46] Fábio Daros de Freitas, Alberto Ferreira De Souza, and Ailson Rosetti de Almeida. A prediction-based portfolio optimization model. 2006.

[47] TWS Chow and CT Leung. Nonlinear autoregressive integrated neural network model for short-term load forecasting. In Generation, Transmission and Distribution, IEE Proceedings, volume 143, pages 500–506. IET, 1996.

[48] Simon Haykin. Kalman Filtering and Neural Networks.2001.

[49] Azadeh, A., Fekri, M., Asadzadeh, S. M., Barazandeh, B., & Barrios, B. A unique mathematical model for maintenance strategies to improve energy flows of the electrical power sector. Energy Exploration & Exploitation, 34(1), 19-41, 2016.

[50] Barazandeh, B., Bastani, K., Rafieisakhaei, M., Kim, S., Kong, Z., & Nussbaum, M. A. (2017). Robust sparse representation-based classification using online sensor data for monitoring manual material handling tasks. IEEE Transactions on Automation Science and Engineering, (99), 1-12.

[51] Sabbagh, R., Ameri, F., & Yoder, R. Thesaurus-guided Text Analytics Technique for Capability-based Classification of Manufacturing Suppliers. Journal of Computing and Information Science in Engineering, 18(3), 031009, 2018.

[52] Sabbagh, R., & Ameri, F. (2018, August). Supplier Clustering Based on Unstructured Manufacturing Capability Data. In ASME 2018 International Design Engineering Technical Conferences and Computers and Information in Engineering Conference (pp. V01BT02A036-V01BT02A036). American Society of Mechanical Engineers.

[53] Fabio D. Freitas, Alberto F. De Souza, Alison R. De Almeida. Prediction-based portfolio optimization model using neural networks, Neuro computing 72 (2009) 2155–2170.